\documentclass[aps,preprint,superscriptaddress,graphicx,longbibliography]{revtex4-1}
\usepackage{graphicx}

\begin{document}


\title{Resistive Switching in Nanodevices}


\author{Hannes Raebiger}
\email[E-mail:]{hannes@ynu.ac.jp}
\affiliation{Centro de Ci\^e{}ncias Naturais e Humanas, Universidade Federal do ABC, Santo Andr\'e, SP, Brazil}
\affiliation{Department of Physics, Yokohama National University, Yokohama, Japan}
\author{Antonio Claudio M. Padilha}
\affiliation{Centro de Ci\^e{}ncias Naturais e Humanas, Universidade Federal do ABC, Santo Andr\'e, SP, Brazil}
\author{Alexandre Reily Rocha}
\affiliation{Instituto de F\'{\i}sica Te\'orica, Universidade Estadual Paulista, S\~ao Paulo, SP, Brazil}
\author{Gustavo M. Dalpian}
\email[E-mail:]{gustavo.dalpian@ufabc.edu.br}
\affiliation{Centro de Ci\^e{}ncias Naturais e Humanas, Universidade Federal do ABC, Santo Andr\'e, SP, Brazil}


\date{\today}

\begin{abstract}
Passing current at given threshold voltages through 
a metal/insulator/metal sandwich structure device
may change its resistive state.
Such resistive switching is unique to nanoscale devices, but 
its underlying physical mechanism remains unknown.
We show that the different resistive states are due to different spontaneously charged states,
characterized by different `band bending' solutions of Poisson's equation.
For an insulator with mainly donor type defects, the low-resistivity state is characterized by 
a negatively charged insulator due to convex band bending,
and the high-resistivity state by a positively charged insulator due to concave band bending;
vice versa for insulators with mainly acceptor type defects.
These multiple solutions coexist only for nanoscale devices and for 
bias voltages limited by the switching threshold values,
where the system charge spontaneously changes
and the system switches to another resistive state.
We outline the general principles how this functionality depends on material properties
and defect abundance of the insulator `storage medium',
and propose a new magnetic memristor device with increased storage capacity.
\end{abstract}


\maketitle

\section{Introduction}

The existence of resistive switching memories, also known as memristors or
Resistive Random Access Memories (ReRAM or RRAM),
was predicted from circuit theory nearly half a century ago~\cite{Chua:1971ce,Chua:1976dm}
and functional devices have been demonstrated within the past decade~\cite{Strukov:2008he,Chua:2011el,Yang:2012gq,Pan:2014gh,Hwang:2015dd}.
The prototypical memristor is composed of a thin slab of TiO$_2$ sandwiched between two metal electrodes~\cite{Szot:2011be},
but memristors have been realized using a wide range of insulator materials as storage medium~\cite{Pan:2014gh,Hwang:2015dd}.
By applying bias voltages in excess of given threshold values, the resistivity of the storage medium can be changed
from a low-resistivity state (LRS) to a high-resistivity state (HRS) and back,
over a large number of cycles~\cite{Pan:2014gh}.
A revolution in information technology is only hindered by
the lack of understanding of the physical principles 
that underlie resistive switching~\cite{Schroeder:2010ht,Yang:2012gq,Pan:2014gh,Yang:2014hq,Hwang:2015dd}.

Thus far, two types of mechanisms have been suggested for this resistive switching.
{\em Electronic mechanisms}
describe the switching as electrons being trapped and un-trapped inside the insulator storage medium,
such that the differently charged systems would exhibit different resistivities~\cite{Chopra:1965kj,Simmons:1967wu,Argall:1968ie}.
{\em Ion drift mechanisms}, on the other hand, attribute the switching to a concerted diffusion of atoms/ions
that leads to a controlled and reproducible formation and breaking of conductive filaments
through the insulator storage medium~\cite{Terabe:2005gk,Szot:2006hg,Waser:2007hw,Yang:2009ia,Kwon:2010be,Gu:2010kg,Kim:2011bs,Kim:2014ux,Shen:2015hp}.
The proposed electronic mechanisms suffer from the {\em voltage-time dilemma}~\cite{Schroeder:2010ht}
because the average lifetimes for the asserted metastable electron-trapping states are much shorter than
typical memristor operation times,
and ensueingly, such a system would spontaneously {\em forget} which resistive state it is in.
Ion drift mechanisms offer arbitrarily long lifetimes for the different resistivity states,
because the corresponding filament structures (formed or broken) are assumed to be stable at room temperature.
This stability, however, implies that the formation and/or breaking of such filaments 
via a concerted motion of ions is a rather slow process
(measured in seconds or minutes~\cite{Szot:2006hg,Yang:2014hq,Kudo:2014fo}),
i.e., much slower than
memristor switching times measured in nanoseconds or less~\cite{Pan:2014gh}.
Even accelerated ion diffusion theories~\cite{Ielmini:2011fn,Menzel:2011io,Ielmini:2012fq,Menzel:2012jf}
predict switching times in the order of $10^{-4}$~s.
Moreover, neither mechanism justifies typical on/off resistivity ratios 
($>10^5$ for TiO$_2$~\cite{Pan:2014gh}, and even much larger for other materials).
For example, assuming the ion drift mechanism for TiO$_2$,
conductivity through observed Magn\'eli phase filaments increases only 
by a factor of $10^2$--$10^3$~\cite{Kim:2011bs,Szot:2011be}.

We present a new electronic mechanism for resistive switching.
The LRS, HRS, and possible intermediate resistivity states (IRS) correspond to different `band bending' solutions of Poisson's equation,
which describes the electrostatic potential and charge stored inside 
the insulator storage medium.
This implies that by controlling the defect abundance and size of the insulator material, 
one can construct a resistive memory using arbitrary insulator materials
sandwiched between suitable metal electrodes.
The present theory of resistive switching is universal, 
and there is no need to invoke different types of switching mechanisms for resistive memories made from different types of materials~\cite{Pan:2014gh},
or specific theories for e.g.\ specific correlated materials~\cite{Dubost:2013ch,Li:2015hi}.
The variationally calculated electrostatics, or band bending solutions presented here are general for any metal/insulator/metal structures (not only memristors),
and as we show, in stark contrast to the typical approximations made in previous theories of resistive switching.
The present theory does not exclude the existence of the formation and breaking of conductive filaments during memristor operation~\cite{%
Terabe:2005gk,Szot:2006hg,Waser:2007hw,Yang:2009ia,Kwon:2010be,Gu:2010kg,Kim:2011bs,Kim:2014ux,Shen:2015hp},
but 
we propose that it has a different role in resistive switching, as was assumed previously.

This work is organized as follows.
First, the electrostatic model for different resistive states is presented, 
followed by a model device and an electrostatic theory for resistive switching.
We then discuss the influence of material parameters on the multiple resistive states, 
and outline the design principles how to craft memristors using arbitrary insulator materials as storage medium,
followed by a discussion of the role of storage media filamentation in resistive switching.
Finally, we propose a new magnetic memristor device with increased storage capacity,
and summarize our results in the Conclusion.
Details of our variational calculations are given in the Appendix.

\section{Electrostatic description}

The key point that is missing in previous switching mechanism models---electronic or ion drift based---%
is that they all have assumed the electrostatic potential $\phi(z)$ within the insulator storage medium to exhibit
either typical metal-insulator {\em band bendings}, or no band bending at all~\cite{Simmons:1967wu,Schroeder:2010ht}
(here $e\phi$
marks the position of the conduction band minimum, see App.~\ref{sec:methods}).
Such band bendings follow approximate (analytic) solutions to the one-dimensional Poisson's equation 
\begin{equation}
\frac{\partial^2 \phi}{\partial z^2} = -\frac{\rho(z)}{\epsilon \epsilon_0}
\label{eq:poi}
\end{equation}
along $z$
perpendicular to the metal-insulator interface,
where $\epsilon$ and $\epsilon_0$ are the relative and vacuum permittivities, respectively,
and the charge density $\rho(z)$ is approximated by a step function.
This approximation for $\rho(z)$ is justified for a semiconductor with a small concentration of (shallow) dopants,
but not necessarily in memristors, where defect concentrations can be 10$^{21}$~cm$^{-3}$ ($=1$~nm$^{-3}$)
or even larger~\cite{Szot:2011be}.
In fact, $\rho$ is a functional that depends on both $z$ and $\phi(z)$, i.e., $\rho = \rho[\phi(z),z]$, 
and must be solved {\em self-consistently} together with $\phi(z)$.
We thus solve Eq.~(\ref{eq:poi}) variationally by minimizing the functional 
\begin{equation}
F[\phi(z)] = \int_a^b \; {\rm d}z \; \left(  \frac{1}{2}\phi'^2 + \phi \frac{\rho}{\epsilon \epsilon_0} \right) \; .
\label{eq:dir}
\end{equation}
Here $a$ and $b$ denote the $z$ coordinates of the two insulator/metal interfaces,
and $\phi'$ denotes the derivative $\phi' = \frac{{\rm d}\phi}{{\rm d}z}$.
$F$ is an energy functional, whose first term $F_1 = \int_a^b  {\rm d}z \, \frac{1}{2}\phi'^2$ gives the potential energy of the electric field over the insulator,
and the second term $F_2 = \int_a^b {\rm d}z \, \phi \frac{\rho}{\epsilon \epsilon_0}$ the interaction energy between the electrostatic potential and charge stored in the device.

\subsection{Model device}

\begin{figure}
\includegraphics{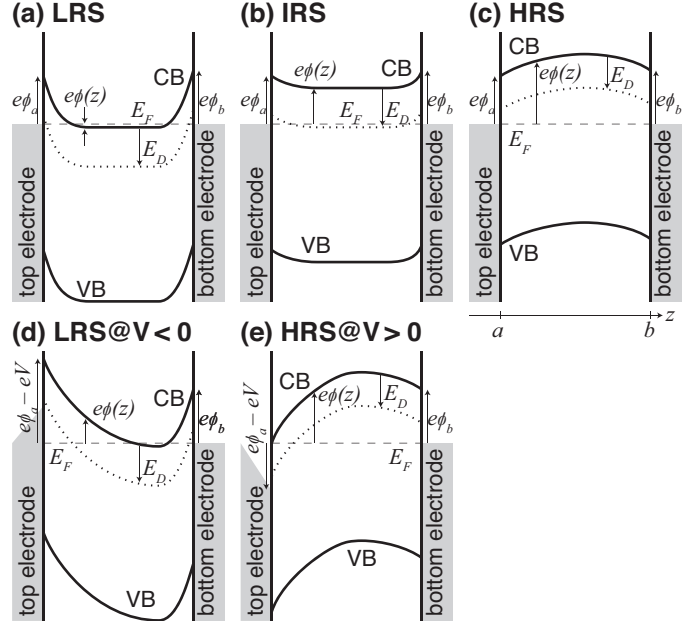}
\caption{Band bending diagrams for different resistive states.
(a) Low resistivity state LRS;
(b) Intermediate resistivity state IRS;
(c) High resistivity state HRS;
(d) LRS at negative bias voltage $V$;
(e) HRS at positive $V$.
The strong bent lines indicate conduction and valence bands CB and VB 
and the dotted line indicates the donor activation energy $E_D$ measured from CB.
The dashed thin line indicates the Fermi level $E_F$.
$V$ applied to the top electrode shifts the band edges at $z=a$ by $eV$ up for negative and down for positive $V$,
as shown in (d) and (e).
}
\end{figure}

We consider a prototype 
memristor that contains donor defects with the activation energy $E_D$ measured from the conduction band (CB).
The insulator storage medium forms Schottky connections with both top and bottom electrodes with Schottky barriers $e\phi_a$ or $e\phi_b$, 
and $E_D$ is assumed smaller than either of these barriers.
Depending on initialization, we find three different families of numerical solutions to Eq.~(\ref{eq:poi})
illustrated by the band bending diagrams shown in Fig.~1~(a)--(c). 
The LRS shown in Fig.~1~(a) is characterized by a convex $\phi(z)$ where most of the CB is filled by electrons
and most donor defects are not ionized;
because of the CB filling, this state exhibits excellent band conductivity.
The other convex $\phi(z)$ solution shown in Fig.~1~(b)
is an IRS, which has the Fermi level ($E_F$) mostly pinned by $E_D$.
Here most donor defects are not ionized, 
but because $E_F$ is very close to $E_D$,
we can expect quite large hopping conductivity.
Note that if $E_D$ is very close to the CB and there is significant overlap of $E_D$ and CB at room temperature,
the IRS becomes engulfed by the LRS, i.e.,
such IRS only appear for deep defects with $E_D$ well inside the band gap.
At the same time, 
in the case of multiple deep levels, there may be multiple IRS's.
Finally, the HRS shown in Fig.~1~(c) exhibits a concave $\phi(z)$ such that all donor defects are ionized,
the CB is completely empty,
and there are no carrier or impurity states nearby $E_F$.
We may expect
poor hopping conductivity at most, i.e., this state is an insulator.
These different conductive mechanisms are supported by recent experimental observations of band conductivity for LRS
and hopping conductivity for HRS~\cite{Goldfarb:2012cd,Chiu:2014fk},
ensued by large on/off resistivity ratios of $10^5$ or more.

\subsection{Resistive switching}

Applying a bias voltage of $V$ to the top electrode shifts the band edges at $z=a$ by $-eV$.
Thus applying a negative bias in the LRS reduces the number of electrons in the CB and ionizes more and more donor defects, as shown in 
Fig.~1~(d), until the system exhibits a transition to the HRS.
The transition from IRS to HRS occurs similarly.
Applying a positive bias to the HRS, on the other hand, when $V > \phi_a - E_D/e$ causes an increasing number of donor levels to be de-ionized,
causing a transition to the IRS, and once $V > \phi_a$ the CB becomes occupied, causing a transition to the LRS.
It is important to note that
while the above discussion holds for a system with mainly donor defects,
the equations are symmetric with respect to charge, and, the above findings hold equally well for systems with acceptor defects
and hole carriers.

\begin{figure}
\includegraphics{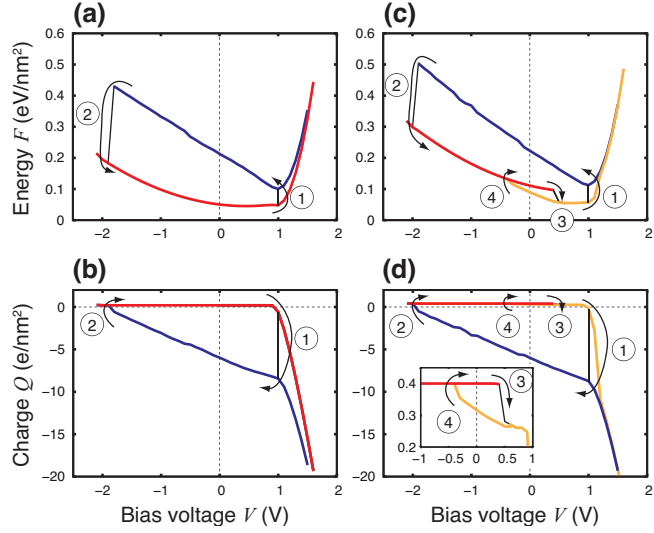}
\caption{
Energy $F$ and charge $Q$ stored in prototype memristor as a function of bias voltage $V$.
(a) and (b) show the energy $F$ and charge $Q$ for defect scenario (I),
a homogeneous distribution of  oxygen vacancies V$_{\rm O}$ with concentration 0.005~nm$^{-3}$.
(c) and (d) show $F$ and $Q$ for defect scenario (II),
a homogeneous distribution of  oxygen vacancies V$_{\rm O}$ with concentration 0.005~nm$^{-3}$
and Ti interstitals Ti$_{\rm i}$ with concentration 0.0025nm$^{-3}$.
$F$ and $Q$ corresponding to HRS, IRS, and LRS are given by red, orange, and blue lines,
and discontinuous jumps at switching between these states is given by the black line segments.
Switching between HRS and LRS is indicated by the arrows labeled by 1 and 2,
and switching between HRS and IRS by the arrows labeled 3 and 4.
The insert in (d) shows a magnification of the HRS and IRS switching.
}
\end{figure}

To illustrate the switching mechanism, 
we calculate 
the energy $F$  and charge $Q$ of our prototype memristor
as a function of bias voltage $V$ at room temperature ($T=$ 300~K), shown in Fig.~2.
The red, orange, and blue lines indicate the HRS, IRS, and LRS,
and the switching from one state to the other is indicated by arrows.
Our prototype memristor is
composed of a 20~nm slab of rutile TiO$_2$ sandwiched between 
metal electrodes that form a Schottky barrier of $e\phi_a = e\phi_b = 1$~eV at both sides.
Here, rutile has a band gap of 3~eV, a relative permittivity of $\epsilon =$ 80,
effective electron masses of $m^*_c =$ 8.5 and 0.4 for heavy and light electrons,
and the effective hole mass of $m^*_v =$ 0.9 (effective masses were calculated from first principles
and are well in agreement with experiment~\cite{Yagi:1996ho}).
We consider two defect scenarios that mimic the oxygen deficiency of TiO$_2$ memristors:
(I) rutile contains oxygen vacancies V$_{\rm O}$ with a concentration of 0.005~nm$^{-3}$ ($=5\times10^{18}$~cm$^{-3}$),
which act as shallow double donors~\cite{Janotti:2010cx};
and
(II) rutile contains 0.005~nm$^{-3}$ of V$_{\rm O}$ together with 0.0025nm$^{-3}$ of Ti interstitals Ti$_{\rm i}$,
which act as a quadruple deep donor with the donor activation energy at $E_D = 0.6$~eV below CB
(see Refs.~\cite{DiValentin:2009fs,Lee:2012hu};
also the Magn\'eli phase filament like structures observed in TiO$_2$ may act as donors~\cite{Padilha:2015vk}).
The memristor energy $F$
for scenario (I) shown in Fig.~2~(a) has two energy branches, 
and for scenario (II) shown in Fig.~2~(c) has three energy branches,
corresponding to the different resistivity states.
These energy branches coexist over a wide range of $V$, and, most importantly, coexist at $V=0$.
The upshot is that once the system is in one of the different resisitivity states, it remains in that state unless a bias in excess of 
the switching thresholds is applied.

\subsection{Charge and electric field inside memristor}

Figs.~2~(b) and (d) show the charge $Q$ stored in the memristor.
Notice that none of the states HRS, IRS, or LRS is charge neutral, i.e., $Q\ne 0$ always.
The insulator storage medium of a memristor thus is spontaneously charged,
such that the HRS and LRS have opposite polarities.
In our example cases, for $V=0$, $\phi$ is symmetric, but in experimental devices, the Schottky barriers of the top and bottom electrodes are not necessarily equal,
and the defect concentration is likely inhomogeneous.
Such asymmetries make the system polar, which combined with charge $Q\ne 0$ stored in the device means that 
it can also act as a nanobattery, 
as observed in recent experiment~\cite{Valov:2013dm}.
When the HRS or IRS is switched to the LRS, there is a flow of electrons into the system,
because the connection with the electrode is essentially ohmic (see Fig.1~(e)).
As $V$ then decreases below the switching threshold, a Schottky barrier is formed again, 
preventing leakage of $Q$ from storage medium to electrodes.
However, as $V$ is further decreased, $Q$ gradually approaches 0, and once the system is depleted from electrons,
it is reset to the HRS; for the LRS, $Q<0$, and for the HRS, $Q>0$.
The switching between HRS and IRS is similar, albeit here for both states $Q>0$.
As $V \longrightarrow \phi_a - E_D/e$, electrons may freely occupy the deep levels $E_D$ due to Ti$_{\rm i}$,
which causes a discontinuous drop in $Q$ magnified in the inset of Fig.~2~(d).
Switching back to HRS occurs as all electrons are depleted from the impurity states.

\begin{figure}
\includegraphics{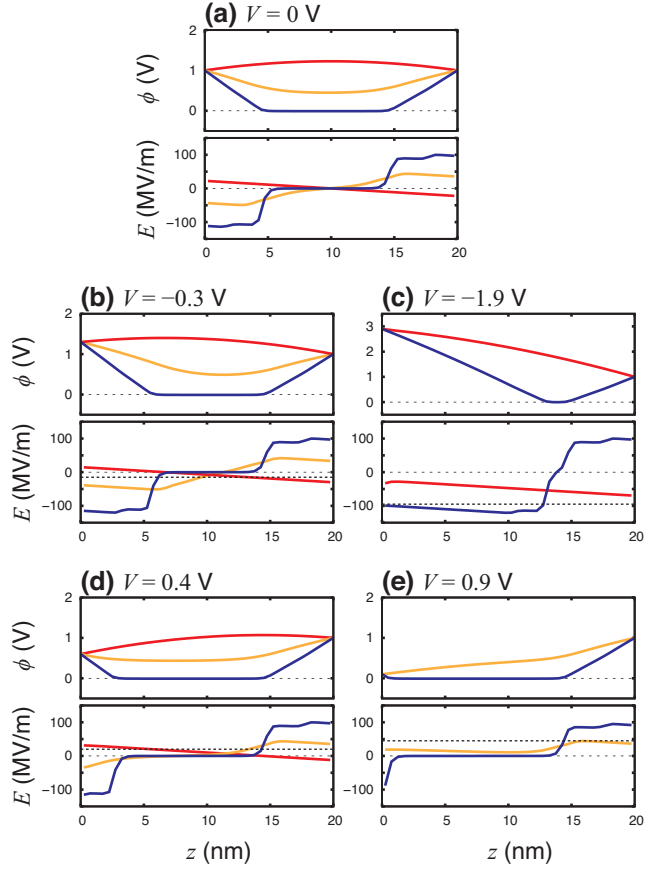}
\caption{Electrostatic potential $\phi$ and electric field $E$ inside memristor storage medium
as a function of $z$ for the 20 nm wide prototype memristor with defect scenario (II)
at different bias voltages $V$.
Notice that $\phi$ is the electrostatic potential for electrons, i.e., the sign convention is opposite to the usual;
with this sign convention $\phi$ illustrates the shape of the storage medium conduction band.
(a) $V=0$; 
(b) $V=-0.3$ V; 
(c) $V=-1.9$ V; 
(d) $V=0.4$ V;
and 
(e) $V=0.9$ V.
Red, orange, and blue lines correspond to HRS, IRS, and LRS.
The horizontal dashed line indicates the constant electric field $E_c = V/L$.
}
\end{figure}

We further illustrate the charging of the memristor by analyzing the electric field $E$ inside the memristor storage medium.
Fig.~3 shows the electrostatic potential $\phi(z)$ and corresponding electric field $E=  \partial \phi / \partial z$ (there is no $-$ sign 
because $\phi$ denotes the electrostatic potential for negative charges;
see App.~\ref{sec:methods})
for our prototype memristor in defect scenario (II)
at zero bias [Fig.~3~(a)] and different bias voltages $V$ immediately below switching threshold values.
At zero bias for the HRS, $E>0$ at the top electrode and $E<0$ at the bottom electrode, and thus any electrons possibly excited to the CB are expelled to either electrode.
This combined with the fact that all donor defects are ionized gives $Q>0$.
For the LRS and IRS, electric field direction has changed and $E<0$ at the top electrode and $E>0$ at the bottom electrode, i.e.,
electrons excited to the CB will be trapped in the central region of the storage medium.
Notice that for the LRS and IRS, most donor defects are not ionized, but nonetheless there are some electrons in the CB of the IRS,
and for the LRS the CB is mostly filled by electrons, and therefore $Q<0$.
Here it is important to recall that at both electrodes, there is a metallic reservoir of electrons,
so any electrons stored inside the memristor (for LRS and IRS) are taken from these reservoirs,
and vice versa for the HRS.

The electric fields inside the memristor storage medium are very large---up to $\pm$100~MV/m---even for zero bias voltage [Fig.~3~(a)].
These internal electric fields are of the same order of magnitude as constant electric fields $E_c = V/L$ 
typically assumed to cause the accelerated diffusion~\cite{Ielmini:2011fn,Menzel:2011io,Menzel:2012jf} of vacancies or impurity atoms
and the ensuing formation or breaking of conductive filaments.
Comparison of $E_c = V/L$ for large bias voltages around switching thresholds [dashed lines in Figs~3~(b)--(e)]
leads to a dilemma:
if an electric field of $E>$50 MV/m would cause oxygen ions from the insulator storage medium to leave their lattice sites (form vacancies) and drift out of the material,
or metal atoms from the electrodes to drift inside the storage medium,
such a process would always take place for the convex band bendings (LRS or IRS) 
close to the electrodes, where the intrinsic electric field is larger than this even at zero bias.
In fact, the band bending and intrinsic electric field at these electrodes 
is similar to what one obtains for a Schottky connection of the same materials~\cite{Monch:1993},
so applying the theory of accelerated diffusion~\cite{Ielmini:2011fn,Menzel:2011io,Menzel:2012jf}
to a simple metal-insulator connection would predict spontaneous
ion diffusion and structural rearrangement within the space charge region.
Thus, in order to understand the possible ion drift mechanisms or filament formation mechanisms in memristors or other metal-insulator systems,
one cannot rely on such simplified approximations of constant electric fields ($E_c \approx \Delta \phi/\Delta z$);
instead, electric fields inside dielectric medium must be correctly evaluated as $E = (-) \nabla \phi$ (see above for sign convention).

Previous electronic switching theories suffer from the voltage-time dilemma~\cite{Schroeder:2010ht},
but this is not the case for the present theory.
This is easily illustrated by either $Q$ or $E$ inside the memristor storage medium.
We must acknowledge that,
in the LRS state, there is always a finite probability $P_{{\rm out}\left(1\right)}$ for one electron to spontaneously tunnel out of the storage medium,
which for previous electronic models would cause the system to spontaneously forget its resistivity.
In the present theory, however, 
LRS is a stationary state, and the probability for the next electron to tunnel out $P_{{\rm out}\left(2\right)}$
is smaller, i.e., $P_{{\rm out}\left(2\right)} < P_{{\rm out}\left(1\right)}$. 
By the same token, the probability for an electron to tunnel back into the storage medium from one of the electrodes $P_{{\rm in}}$
is larger than $P_{{\rm out}\left(2\right)}$,
i.e., the metastable state with one electron removed from the LRS tends to relax back into LRS rather than to decay into HRS.
The same can be concluded based on the internal electric field in the storage medium:
Figs 3~(b) and (d) show the electric fields for small bias voltages below switching thresholds.
For LRS and IRS $E<0$ at the top electrode and $E>0$ at the bottom electrode, which maintains $Q<0$,
and vice versa for the HRS.
After such "read" operations at zero bias, the intrinsic electric fields drive the system towards the stationary states shown in Fig.~3~(a).
The charge difference between HRS and IRS in our prototype, however, is smaller than 1 electron per nm$^2$,
so the occasional tunneling of just one electron out of the system can cause the system to {\em forget} its state.
Thus, for a reliable memristor, the difference in $Q$ between the different resistivity states should be as large as possible.
This charge difference can be maximized by choosing storage medium materials with a large electron (or hole) effective mass.

\section{Memristor materials}

\begin{figure}
\includegraphics{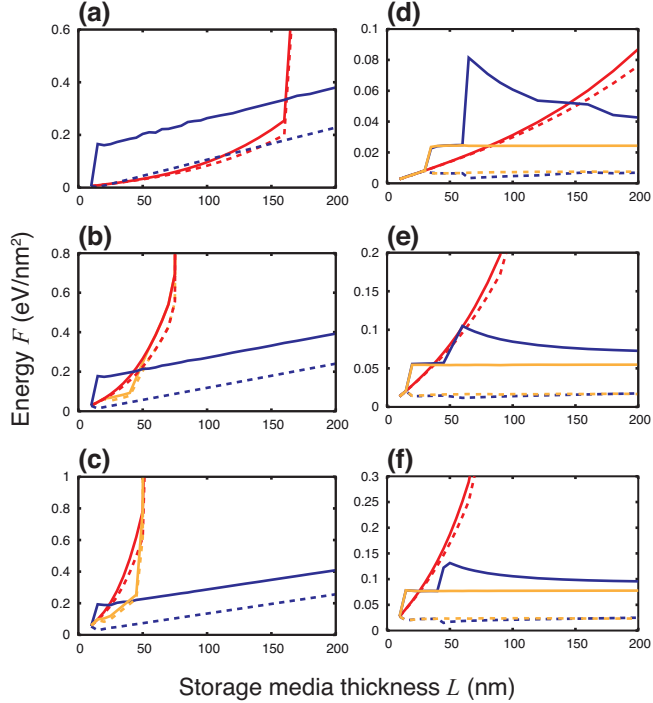}
\caption{Memristor energies $F$ as a function of storage medium thickness $L$.
(a), (b), and (c) show $F$ and its component $F_2$ for rutile TiO$_2$
with
V$_{\rm O}$ at the concentrations of (a) 0.001~nm$^{-3}$,
(b) 0.005~nm$^{-3}$, and (c) 0.01~nm$^{-3}$
and a correlated concentration of Ti$_{\rm i}$ ([${\rm Ti_i}] = \frac{1}{2} [{\rm V_O}]$).
(d), (e), and (f) show $F$ and $F_2$ for ZnO with
Cr impurities at the concentrations of (d) 0.0001~nm$^{-3}$,
(e) 0.0005~nm$^{-3}$, and (f) 0.001~nm$^{-3}$.
The red, orange, and blue solid lines show HRS, IRS, and LRS,
and the dashed lines show the corresponding $F_2$.
}
\end{figure}

The properties of the different resistivity states depend on the memristor material parameters,
defect abundances, and storage medium thickness $L$.
An obvious prerequisite for resistive switching is the coexistence of the different resistive states.
We investigate this coexistence by evaluating the energy functional $F$ and its components $F_1$ and $F_2$
as a function of $L$
for HRS, IRS, and LRS in two completely different materials, shown in
Fig.~4.
$F$ and $F_2$ are shown for rutile TiO$_2$ with
V$_{\rm O}$ at concentrations of 0.001~nm$^{-3}$ [Fig.~4~(a)],
0.005~nm$^{-3}$ [Fig.~4~(b)], and 0.01~nm$^{-3}$ [Fig.~4~(c)]
and a correlated concentration of Ti$_{\rm i}$ ([${\rm Ti_i}] = \frac{1}{2} [{\rm V_O}]$).
The same are shown for ZnO with 
Cr impurities at the concentrations of 0.0001~nm$^{-3}$ [Fig.~4~(d)],
0.0005~nm$^{-3}$ [Fig.~4~(e)], and 0.001~nm$^{-3}$ [Fig.~4~(f)].
Cr in ZnO is a deep impurity with two donor levels at 1.0 and 2.6~eV below the conduction band~\cite{Raebiger:2009bq};
ZnO has a band gap of 3.4~eV, effective masses $m^*_c =$ 0.28 and $m^*_v =$ 0.6, and a relative permittivity of $\sim$ 10.
$F$ for the HRS increases rapidly as $L$ increases,
mainly due to the similar increase of $F_2$.
At $L$ larger than some threshold value, $F$ for HRS becomes exceedingly large ($F^{\rm HRS} >> F^{\rm LRS}$),
i.e., HRS becomes unstable.
This sets a maximum size for the memristor, which for TiO$_2$ and ZnO is estimated to be some tens of nm
depending on the defect concentration.

The superlinear behavior of $F_2$ described above arises as follows.
$F_2$ is proportional to $\rho$ multiplied by $\phi$ integrated over the length $L$ of the memristor.
For the HRS, all defects are ionized so $\rho$ is constant, proportional to defect concentration $N_D$.
At the same time, for HRS, $\phi$ is increasingly bent upwards as $L$ increases, which leads to the superlinear behavior of $F_2$.
The curvature of $\phi$ is inversely proportional to the relative permittivity, and hence, $F_2$ increases much more rapidly
in ZnO than TiO$_2$ at similar defect concentrations (cf. panels (a) and (f) in Fig.~4).
For small $L$ (around 10 nm or smaller), HRS, IRS, and LRS are indistinguishable,
because there is no band bending.
This is because the curvature of $\phi$ is proportional to defect concentration.
It follows that in ZnO, where $F$ for the HRS is reasonably small only for very small defect concentrations,
the IRS and LRS with convex band bendings only emerge at rather large $L$.
The above discussion illustrates the interplay of relative permittivities and defect concentrations,
and serves as a guideline to design memristors using arbitrary insulator materials as storage medium.

\subsection{The role of filamentation in resistive switching}

Filament formation inside the memristor storage medium has been observed in various devices~\cite{%
Terabe:2005gk,Szot:2006hg,Waser:2007hw,Yang:2009ia,Kwon:2010be,Kim:2011bs,Kim:2014ux,Shen:2015hp}.
For example, conical regions of oxygen deficient Magn\'eli phases have been observed in TiO$_2$ memristors,
and chain like structures of metal impurity atoms have been observed in Conductive Bridging Random Access Memory (CBRAM) 
devices~\cite{Yang:2012cz,Celano:2014bw}.
It is clear that many memristor devices contain such filament like structures,
but their role in resistive switching is not so obvious.
While voltage controlled filament formation and manipulation has been demonstrated,
the time scales for such concerted ionic motion are much slower (seconds or minutes) than typical resistive switching times (nanoseconds or less).
A recent in-situ transmission electron microscopy (TEM) study of a memristor device
shows that resistive switching is accompanied by the appearance and disappearance of 
a dark region in TEM images ensuing the switching to LRS and to HRS, respectively,
which was interpreted as the formation and erasure of conductive filaments~\cite{Kudo:2014fo}.
However, these changes appear in the TEM images {\em after} the switching, over a time period of $\sim0.15$ seconds,
which would suggest that the contrast change in TEM is a consequence of the switching, rather than its cause.

One of the present authors has shown that
quasi-1D metal impurity chains inside a semiconductor medium
behave akin to a large concentration of deep impurities 
the semiconductor~\cite{Raebiger:2014jd}, rather than a metallic inclusion.
These 1D chains exhibit multiple (deep) donor or acceptor transitions, shifted from similar transitions observed for isolated impurities.
We have also shown that the TiO Magn\'eli phases exhibit a semiconductor band structure with an intermediate band~\cite{Padilha:2014fp,Padilha:2015vk}, 
which essentially may behave like a deep donor state.
Thus, instead of the filaments behaving like a metallic lead connecting the electrodes,
we propose that their role is electronic, i.e., they behave as donor or acceptor type defects, i.e.,
in the LRS, the conductivity is not through the electronic states of the filament, but the insulator conduction band.

\subsection{Magnetic memristor}

The above case study of a memristor based on Cr-doped ZnO discussed above leads to an intriguing extra feature.
Transition metal doped insulators may exhibit carrier-mediated ferromagnetism~\cite{Dietl:2000tw,Dietl:2014bk}.
For Cr-doped ZnO (or In$_2$O$_3$)~\cite{Lany:2008bf,Raebiger:2008dt} the mechanism of magnetism is such that
the ferromagnetic interaction is switched on by degenerate electron doping, i.e., by filling the CB by carrier electrons.
Thus, the LRS of a memristor using Cr-doped ZnO or In$_2$O$_3$ would be spontaneously magnetized.
The polarity of the storage medium magnetization could be controlled by a ferromagnetic top electrode,
i.e., the storage medium is assumed to adopt the polarity of the top electrode at the instant of switching.
Such a magnetic memristor has four spin configurations in the LRS, as shown in Fig.~5.
For an asymmetric memristor, all four spin configurations are expected to have different resistivities
due to spin-dependent transport (see e.g. Ref.~\cite{Prinz:1999to}),
leading to increased storage capacity of the memristor device.
The spin-dependent transport could be enhanced by having a half-metallic ferromagnet~\cite{DEGROOT:1983vn,Pickett:2001wn}
as one of the metal contacts,
such as e.g. the GdN/GaN interface.~\cite{Kagawa:2014iy}

\begin{figure}
\includegraphics{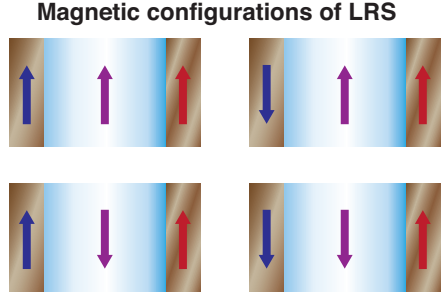}
\caption{Spin configurations of magnetic memristor.
The polarity of the bottom electrode is assumed to be fixed (red arrow),
whereas the top electrode polarity (blue arrow) is assumed to be easily reversible (writable).
The storage medium polarity (purple arrow) can be controlled by the polarity of the top electrode at the time of switching.
}
\end{figure}

\section{Conclusion}

We present an electronic mechanism of resistive switching for nanodevices.
This mechanism is demonstrated by self-consistent calculations of 
the electrostatic potential $\phi(z)$ from Poisson's equation for prototype 
resistive memory devices, 
composed of an insulator slab
sandwiched between metallic electrodes that form Schottky barriers with the insulator.
The two key ingredients for resistive switching are 
(i) the insulator must be abundant with donor or acceptor type defects 
and
(ii) the insulator slab thickness has to be less than some tens or hundreds of nanometers;
the maximal thickness depends on the insulator properties and defect abundance.
When both conditions are met, the system exhibits multiple band bending solutions,
which may correspond to the high resistivity state, where the insulator is {\em completely depleted} of charge carriers,
the low resistivity state where the insulator valence or conduction band contains a maximal amount of charge carriers,
or intermediate resistivity states, where the Fermi level is pinned around deep impurity levels inside the gap.
The {\em right amount} of defects can be obtained during initial electroforming, 
which has been demonstrated to form various kinds of defects in the insulator storage medium~\cite{Pan:2014gh,Yang:2014hq,Szot:2011be,Kwon:2010be,Kim:2011bs}.
These different $\phi$ are stationary states, robust against small perturbations, 
and stochastic tunneling of electrons into or out of the insulator storage medium cannot change the resistive state of the device---%
switching occurs only once the bias voltage exceeds given threshold values.
Due to the electronic nature of the switching, it may be possible to switch resistive states also optically,
provided the device is sufficiently optically transparent.
This electronic nature also explains the fast switching times.
Finally, 
we propose a magnetic memristor, which
could allow simultaneous control of both $\phi$ and magnetic states of the system, and increase the storage capacity of the memristor.

\appendix
\section{Self-consistent calculations}
\label{sec:methods}

Let us consider an insulator material sandwiched between two metal electrodes at $z = a$ and $z=b$,
which form Schottky contacts with $n$ type Schottky barriers $e\phi_a$ and $e\phi_b$,
i.e., a one-dimensional Poisson equation with the Dirichlet boundaries $\phi(a) = \phi_a$ and $\phi(b) = \phi_b$.
$\phi(z)$ is the electrostatic potential of an electron (charge $-e$)
chosen such that $e\phi$
marks the position of the conduction band minimum with respect to electron chemical potential
(Fermi level) fixed at $e\phi = 0$. 
To find the $\phi$ that minimize Eq.~(\ref{eq:dir}), we solve the Euler equation
\begin{equation}
\frac{{\rm d}^2\phi}{{\rm d}z^2} = -\frac{\rho}{\epsilon \epsilon_0} + \frac{\phi}{\epsilon \epsilon_0} \frac{\partial \rho}{\partial \phi} \; 
\label{eq:euler}
\end{equation}
self-consistently with the charge density $\rho$.

The charge density in Eq.~(\ref{eq:euler}) is
\begin{equation}
\rho = - n_c  + p_v + N_{\rm D} (q_{\rm D,0} - n_{\rm D}) \; .  
\label{eq:rho}
\end{equation}
$\rho$ is composed of the densities of 
electrons in the conduction band ($n_c$),
holes in the valence band ($p_v$), and the concentration of
donor or acceptor defects ($N_{\rm D}$) multiplied by the charge of that defect $q_{\rm D,0}$ when all its gap levels are empty
minus the number of electrons occupying the defect induced levels ($n_{\rm D}$).
Electron and hole densities in conduction and valence bands, $n_c$ and $p_v$
are calculated as
\begin{eqnarray}
&&n_c = \int_{e\phi}^\infty \; {\rm d}\varepsilon \; \frac{g_c(\varepsilon)}{\exp(\varepsilon/kT) + 1} \\
&&p_v = \int_{-\infty}^{e\phi-\varepsilon_{\rm gap}} \; {\rm d}\varepsilon \; \frac{g_v(\varepsilon)}{\exp(-\varepsilon/kT) + 1} \; .
\end{eqnarray}
$\varepsilon$ is the electron quasiparticle energy,
and $g_v$ and $g_c$ are effective densities of states of the valence and conduction bands given by
\begin{eqnarray}
&&g_c(\varepsilon) = \frac{{m^*_c}^{3/2}}{\hbar^3 \pi^2} \sqrt{2(\varepsilon - e\phi)} \\
&&g_v(\varepsilon) = \frac{{m^*_v}^{3/2}}{\hbar^3 \pi^2} \sqrt{2(e\phi - \varepsilon_{\rm gap} -\varepsilon)} \; .
\end{eqnarray}
Here $m^*_c$ and $m^*_v$ denote the effective masses of the conduction and valence bands,
and $\hbar$ is the Planck constant.
The number of electrons occupying donor or acceptor levels in the gap is given by
\begin{equation}
n_{\rm D} =  \frac{\sum N_j \exp(-\varepsilon_j / kT)}{\sum \exp(-\varepsilon_j / kT)} \; ,
\end{equation}
where $\varepsilon_j$ and $N_j$ are the energy and number of electrons in state $j$,
and the summations are taken over all electronic states within the gap.

A self-consistent solution is obtained as follows.
We start by evaluating $\rho$ (Eq.~(\ref{eq:rho})) for some trial $\phi$. 
Then we solve Eq.~(\ref{eq:euler}) numerically, using the finite difference approximation, to obtain a new $\tilde{\phi}$.
Using Eq.~(\ref{eq:rho}) again, we construct a new $\tilde{\rho}$.
We continue such iterations until 
the energy given by Eq.~(\ref{eq:dir}) converges.
Starting from different trial $\phi$ can lead to completely different solutions.
\begin{acknowledgements}

This work was funded by grants 2013/22577-8, 2011/21719-8, 2010/16202-3, 2011/19924-2, and 2015/05830-7
from the S\~ao Paulo Research Foundation FAPESP 
and also from the Brazilian National Council for Scientific and Technological Development CNPq.

\end{acknowledgements}


%



\clearpage

\section*{Supproting Information}

The resistive switching is demonstrated by 
a movie illustrating the evolution of the electrostatic potential $\phi(z)$ during voltage sweeps.

\section*{ToC Entry}

{\bf Sandwich structure nanodevice has different band bendings that correspond to different resistivities.}
For high resistivity state, the insulator storage medium is completely depleted of charge carriers;
for low resistivity state, storage medium contains maximal amount of charge carriers.
Bias voltage shifts top electrode Schottky barrier, and at given threshold bias, system switches resistivity.

\includegraphics{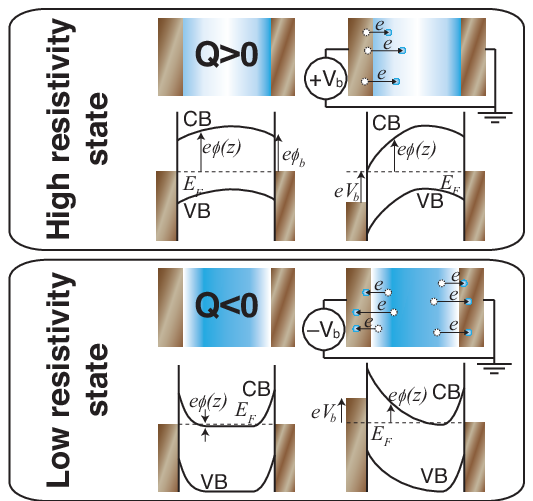}

\end{document}